\documentclass[aps,twocolumn,pre,superscriptaddress,showpacs,reprint]{revtex4-1}
\usepackage{times}
\usepackage{subfigure}
\usepackage{xspace}
\usepackage{graphicx}
\usepackage{dcolumn}
\usepackage{amsmath, amsthm, amssymb}

\newcommand{\ucsdboth}{Department of Physics and Center for Theoretical Biological Physics, University of California,
                          San Diego, La Jolla CA 92093}

\newcommand{\eq}{Eq.~}
\newcommand{\tm}{\ensuremath{t_{\textrm{MFPT}}}}
\newcommand{\eqs}{Eqs.~}
\newcommand{\fig}{Fig.~}

\newcommand{\vb}[1]{ {\bf #1}}

\begin{document}

\title{Velocity alignment leads to high persistence in confined cells}
\author{Brian~A.~Camley}
\affiliation{\ucsdboth}
\author{Wouter-Jan Rappel}
\affiliation{\ucsdboth}

\begin{abstract}
Many cell types display random motility on two-dimensional substrates, but crawl persistently in a single direction when confined in a microchannel or on an adhesive micropattern.  Does this imply that the motility mechanism of confined cells is fundamentally different from that of unconfined cells?
We argue that both free- and confined- cell migration may be described by a generic model of cells as ``velocity aligning" active Brownian particles previously proposed to solve a completely separate problem in collective cell migration.    Our model can be mapped to a diffusive escape over a barrier and analytically solved to determine the cell's orientation distribution and repolarization rate. In quasi-one-dimensional confinement,  velocity-aligning cells maintain their direction for times that can be exponentially larger than their persistence time in the absence of confinement.  Our results suggest an important new connection between single- and collective- cell migration: high persistence in confined cells corresponds with fast alignment of velocity to cell-cell forces.   
\end{abstract}

\maketitle

\section{Introduction}
In traveling through the body, cells move through profoundly complex environments, interacting with nearby cells and extracellular matrix.  By contrast, experiments on cell motility primarily study cells on two-dimensional homogeneous substrates.  Understanding how cells alter their motility in response to confinement is an ongoing and crucial problem in biology that may be relevant to cancer cell migration, where cancer cells may follow preexisting structures or push through tissue \cite{wirtz2011physics,irimia2009spontaneous}.  Confinement of cells may determine invasiveness \cite{ulrich2010probing} as well as changing cell motility modes \cite{friedl2003tumour}.  To study confinement and adhesion in cell motility, controllable {\em in vitro} environments have been developed, including adhesive micropatterned substrates \cite{doyleyamada,chang2013guidance,fraley2012dimensional} and microchannels \cite{irimia2009spontaneous,balzer2012physical,pathak2012independent}.  Cells in confinement can have strikingly different motion than on a substrate \cite{doyleyamada,camley2013periodic,fraley2012dimensional}.  In particular, confinement can change the persistence of the cell's orientation: many cell types undergo persistent directed migration on narrow micropatterned adhesive stripes \cite{doyleyamada} and within small microchannels \cite{irimia2009spontaneous}, even though they undergo primarily random motility on two-dimensional substrates.  

Do these remarkable changes in the character of cell persistence necessarily require free and confined migration to have different biophysical mechanisms?  No!  We show that a single minimal model of cell motility originally proposed to describe a completely independent situation in collective cell migration describes both regimes.  Our model has the critical benefit of an analytic solution, permitting simple comparison to experiment.  This minimal approach has been historically crucial in cell motility, where persistent random walks, run-and-tumble dynamics, and generalizations \cite{selmeczi2005cell,campos2010persistent,cates2012diffusive} have been used to successfully describe and fit experimental eukaryotic and bacterial cell trajectories.  Generic models are also favored here, as different cell types display similar behavior.  

Confined cells are also an important example of boundaries altering the behavior of active matter \cite{marchetti2013hydrodynamics}, an area of physics that is only beginning to be understood.  Boundaries may induce spontaneous circulation \cite{woodhouse2012spontaneous}, rectification \cite{wan2008rectification,angelani2011active}, driving of gears \cite{sokolov2010swimming}, aggregation at edges \cite{elgeti2013wall,fily2014dynamics,lee2013active}, and other complex dynamics \cite{pototsky2012active,nash2010run,hennes2014self,yang2014aggregation}.

Recently, ``velocity aligning'' (VA) active particles and related models have been proposed to describe collective motility of keratocytes \cite{szabo2006phase}, endothelial monolayers \cite{szabo2010collective}, wound healing \cite{basan2013alignment}, and more \cite{henkes2011active,doxzen2013guidance,kabla2012collective,camleyprm}.  VA creates collective motion even though a cell does not ``sense'' its neighbor's orientation.  We use a minimal VA model of a cell as an active Brownian particle that aligns its polarity (internal compass) with its velocity \cite{szabo2006phase}.  We show analytically that under strong confinement, velocity alignment significantly increases a cell's persistence.  Surprisingly, the polarity of a confined VA cell can be mapped to a diffusion-over-a-barrier problem.  This minimal model demonstrates that physical confinement can dramatically alter a cell's type of motility without requiring a different mechanism for free- and confined- cell migration.  

\section{Velocity alignment increases persistence}

In the overdamped active Brownian particle model we study here \cite{szabo2006phase}, a velocity-aligning cell's position $\vb{\tilde{r}} = (\tilde{x},\tilde{y})$ and polarity $\vb{\hat{p}} = (\cos\theta,\sin\theta)$ follow
$\partial_{\tilde{t}} \vb{\tilde{r}} = v_0 \vb{\hat{p}} + \mu \vb{F}$ and 
$\partial_{\tilde{t}} \theta = \frac{1}{T} \textrm{arcsin}\left[ \left(\vb{\hat{p}}\times\vb{\hat{v}}\right)_z \right] + \zeta(\tilde{t})$,
where $\vb{\hat{v}} = (\partial_{\tilde t} \vb{\tilde{r}})/|\partial_{\tilde t} \vb{\tilde{r}}|$ is the cell's velocity direction, $\mu$ is the cell's mobility, $v_0$ is the cell's speed in the absence of external force, and $\vb{F}$ the external force.  The noise $\zeta(\tilde{t})$ has variance $\langle \zeta(\tilde{t}) \zeta(\tilde{t}') \rangle = 2 P^{-1} \delta(\tilde{t}-\tilde{t}')$, where $P$ is the cell's angular persistence time.  The aligning term $T^{-1} \textrm{arcsin}\left[ (\hat{\vb{p}}\times\hat{\vb{v}})_z \right] = T^{-1} \textrm{arcsin}\left[ \cos \theta \hat{v}_y - \sin \theta \hat{v}_x \right]$ makes the polarity direction $\theta$ relax to the velocity direction, $\theta_{\vb{v}} = \textrm{arg} \, \vb{v}$ with a time scale $T$; it is a periodic extension of $-T^{-1}\left(\theta-\theta_{\vb{v}}\right)$ \cite{henkes2011active}.  We describe a single velocity-aligning cell confined within a harmonic potential of stiffness $k_s$ in the $x$ direction.  We rescale lengths and times as $t = \tilde{t}/P$ and $x = \tilde{x} / (v_0 P)$.  In these units,
\begin{align}
\label{eq:r}
\partial_t \vb{r} &= \vb{\hat{p}} - \kappa x \vb{\hat{x}} \\
\partial_t \theta &= \frac{1}{\tau} \textrm{arcsin}\left[ \left(\vb{\hat{p}}\times\vb{\hat{v}}\right)_z \right] + \xi(t) \label{eq:theta}.
\end{align}
$\kappa$ is the unitless measure of the strength of the cell's confinement, $\kappa = \mu k_s P$. $\tau = T/P$ is the ratio of the time required to align the cell's polarity to its velocity to the time scale for the cell's orientation to randomly reorient; a smaller value of $\tau$ implies a more effective aligning mechanism.  The Gaussian Langevin noise $\xi(t)$ has zero mean and variance $\langle \xi(t) \xi(t') \rangle = 2 \delta(t-t')$.  

\begin{figure}[ht!]
\includegraphics[width=80mm]{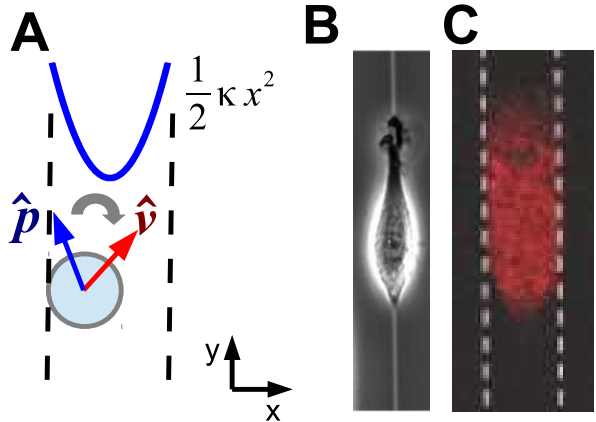}
\caption{A: Velocity-aligning cell model originally proposed by Szabo et al. \cite{szabo2006phase}.  The cell is propelled along its internal polarity $\vb{\hat{p}}$; this polarity aligns itself to the direction of the cell's velocity $\vb{v}.$  The cell moves under the combination of its motility and the force from confinement (\eqs \ref{eq:r}-\ref{eq:theta}).  Dashed lines indicate the stall point $x = \pm \kappa^{-1}$ that the cell cannot cross.  B: Fibroblast on narrow adhesive microstripe; adapted from \cite{doyleyamada}.  C: Human breast adenocarcinoma in microchannel; adapted from \cite{irimia2009spontaneous}.}
\label{fig:illus}
\end{figure}

\begin{figure}[ht!]
\includegraphics[width=90mm]{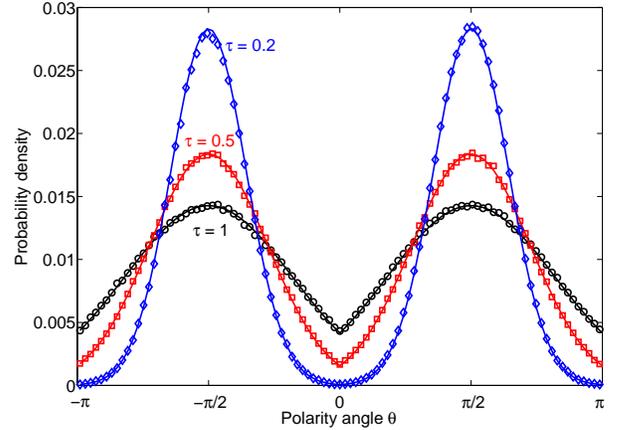}
\caption{Probability distribution of cell polarity angles $\theta$ (modulo $2 \pi$) are given by the Boltzmann-like distribution $p(\theta) \sim \textrm{exp}\left[-W(\theta)/\tau\right]$.  Symbols are Brownian dynamics simulations of our full model (\eqs \ref{eq:r}-\ref{eq:theta}), lines are the Boltzmann distribution.  Simulations performed with $\kappa = 1000$, $\Delta t = 10^{-4}$, and a simulation time of $10^6$.  }
\label{fig:distribution}
\end{figure}

We can solve the $x$ component of our position equation, $\partial_t x = \cos \theta - \kappa x$ for $x(t)$ as a functional of the angle $\theta(t)$, $x(t) = \int_{-\infty}^{t} dt' e^{-\kappa (t-t')} \cos \theta(t')$. We model strongly confined cells (\fig \ref{fig:illus}), and thus assume $\kappa \gg 1$.  In this limit, $e^{-\kappa(t-t')} \to \frac{1}{\kappa} \delta(t-t')$ and $\kappa x(t) \approx \cos \theta(t)$, i.e. the cell quickly crawls to its stall point where $v_x = 0$.  We make the approximation that $v_x = 0$, $\hat{\vb{v}} = \frac{\sin \theta}{|\sin \theta|} \hat{y}$.  Our equation for $\theta$ thus becomes
$\partial_t \theta = \frac{1}{\tau} \textrm{arcsin} \left[ \frac{\cos \theta \sin \theta}{|\sin \theta|}\right] + \xi(t)$.  
Using standard trigonometric identities, we can show that for $\theta \in [-\pi,\pi]$, $\textrm{arcsin} \left[ \frac{\cos \theta \sin\theta}{|\sin\theta|} \right] = -\partial_\theta W(\theta)$, where
\begin{equation}
W(\theta) = -\frac{\pi}{2} |\theta| + \frac{1}{2} \theta^2
\label{eq:W}
\end{equation}
For strong confinement, $\theta$ thus follows
\begin{equation}
\partial_t \theta = -\tau^{-1} \partial_\theta W(\theta) + \xi(t) \label{eq:well}
\end{equation}
where now we interpret $W(\theta)$ as the periodic extension of its definition in \eq \ref{eq:W}.  

\eq \ref{eq:well} provides us with a great deal of insight into the dynamics of the cell's polarity $\theta$: it is precisely the dynamics of a Brownian particle with unit temperature diffusing passively in a potential $\tau^{-1} W(\theta)$.  The potential $W(\theta)$ has minima at $\pm \pi/2$, when the cell polarity is aligned in the $\pm \vb{\hat{y}}$ direction (along the channel).  The velocity alignment parameter $\tau$ acts as an effective temperature: as $\tau \to 0$, $\theta$ becomes increasingly localized to the minima of $W(\theta)$, e.g. the $\pm \vb{\hat{y}}$ directions.  The distribution of angles (modulo $2\pi$) is just the standard Boltzmann distribution, $p(\theta) \sim \textrm{exp}\left[-W(\theta)/\tau\right]$.  We show these distributions, and corresponding ones from direct Brownian dynamics simulation of \eqs \ref{eq:r}-\ref{eq:theta} in \fig \ref{fig:distribution}.  

We can use \eq \ref{eq:well} to compute the rate of spontaneous repolarization of our cells.  In order for a cell to change its direction $\theta$ from $\pi/2$ to $-\pi/2$, it must cross the barrier at $\theta = 0$.  This rate will decrease strongly with decreasing $\tau$, leading to increased persistence of cell motion.  Determining the rate of escape over a barrier is a classic problem; in one dimension, the mean first-passage times can be found exactly up to quadrature via the Smoluchowski equation corresponding to \eq \ref{eq:well} and its adjoint \cite{hanggi1990reaction,zwanzig2001nonequilibrium,kramers1940brownian}.  The mean first passage time from $\theta = \pi/2$ to absorbing boundaries at $\theta = 0,\pi$ is given by 
\begin{align} \label{eq:mfpt} \tm = 
\frac{1}{2}\int_0^\pi dz&\, e^{W(z)/\tau} \, \int_{0}^z dy\, e^{-W(y)/\tau} \\
\nonumber &- \int_0^{\pi/2} dz \, e^{W(z)/\tau} \int_0^z dy \, e^{-W(y)/\tau},
\end{align}
using the symmetry of $W(z)$ about $\pi/2$.  This integral may be evaluated exactly using Mathematica,
\begin{equation}
\tm = \frac{\pi^2}{8} \,_2F_2\left(1,1; \frac{3}{2},2; \frac{\pi^2}{8 \tau}\right) \label{eq:mfpt_exact}
\end{equation}
where $\,_2F_2(a_1,a_2;b_1,b_2;z)$ is the generalized hypergeometric function.  A more convenient and intuitive form that is asymptotically correct in the limit of $\tau\to 0$ can be found by applying the method of steepest descent to the integrals in \eq \ref{eq:mfpt},
\begin{equation}
\tm \approx \sqrt{\frac{2}{\pi}} \tau^{3/2} \textrm{exp}\left(\frac{\pi^2}{8\tau}\right) \label{eq:steepest}
\end{equation}
As expected from our analogy with diffusion over a barrier, the time to repolarize increases exponentially in $1/\tau$: the faster the cell aligns to its local velocity, the longer it takes to turn around.  We compare \eq \ref{eq:mfpt_exact} and \eq \ref{eq:steepest} with mean first passage times (escape rates) observed in Brownian dynamics simulations of \eqs \ref{eq:r}-\ref{eq:theta} in \fig \ref{fig:rates}, and find excellent agreement \cite{footnote_factor_of_two}; see Appendix \ref{app:turnaround} for a discussion of how we measure turnaround times.  
\begin{figure}[ht!]
\includegraphics[width=90mm]{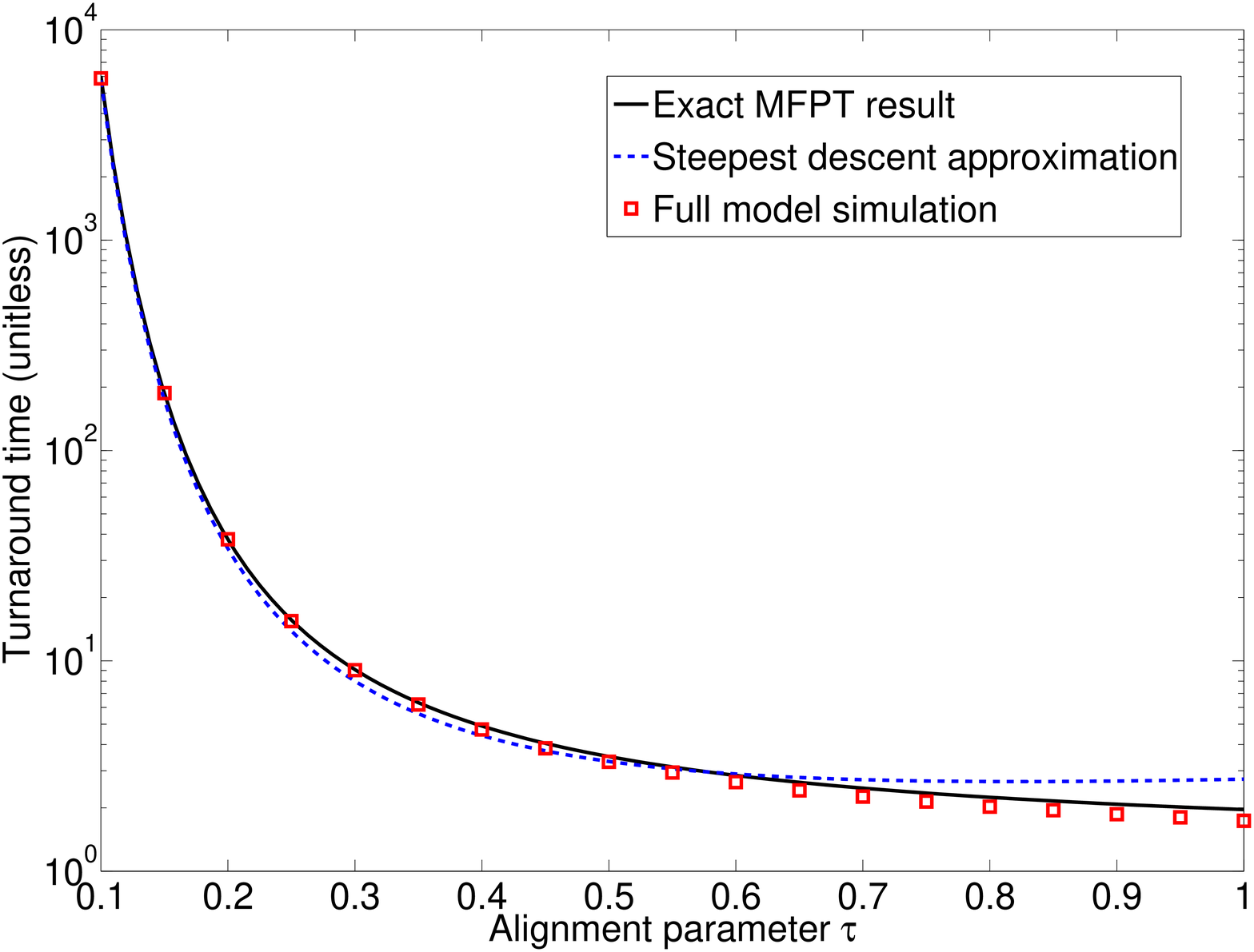}
\includegraphics[width=90mm]{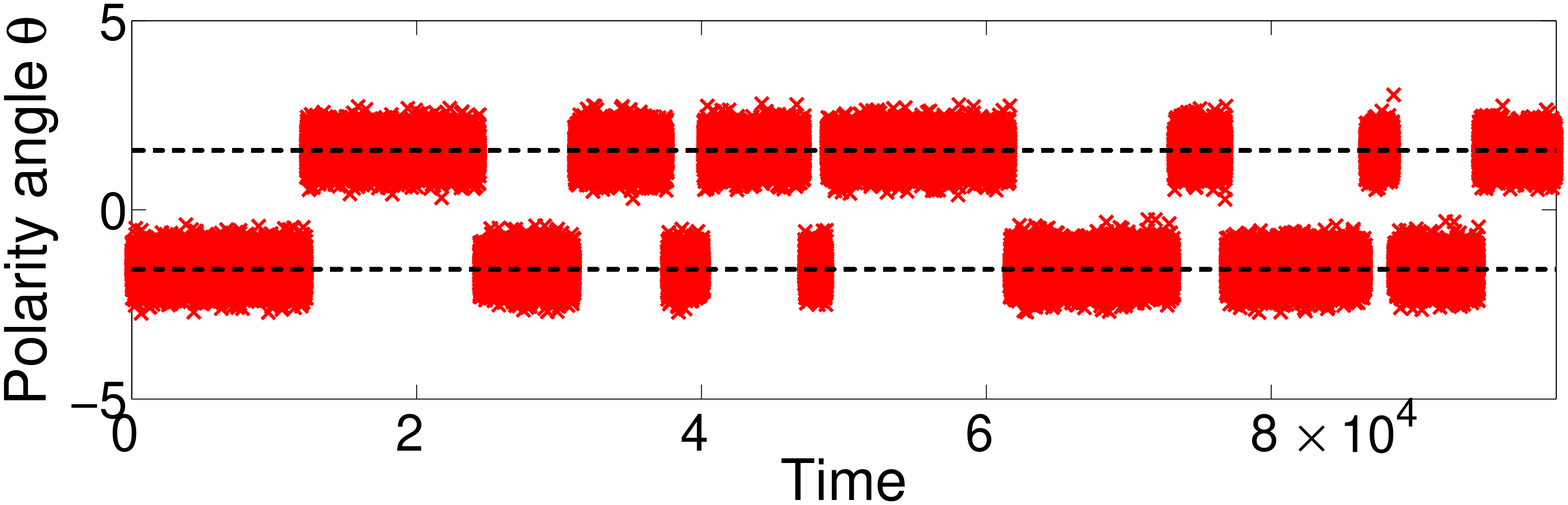}
\caption{When the alignment parameter $\tau$ is small, the time required for the cell to repolarize becomes extremely large.  TOP: Repolarization times from simulation of \eqs \ref{eq:r}-\ref{eq:theta} (symbols) and theory; MFPT result is \eq \ref{eq:mfpt}, steepest-descent result is \eq \ref{eq:steepest}.  BOTTOM: Typical time trace for $\theta$ (modulo $2\pi$) for $\tau = 0.1$.  Dashed lines indicate $\pm \pi/2$.  Simulations performed with $\kappa = 1000$, $\Delta t = 10^{-4}$, and a simulation time of $10^6$.}  
\label{fig:rates}
\end{figure}

While our MFPT theory very accurately predicts the Brownian dynamics simulation, there is some discrepancy at large values of $\tau$.  Here, the MFPT theory overestimates the time required to escape the potential well.  This occurs because we have assumed that the escape rate for the entire well is effectively the same as the escape rate for the center, $\pi/2$.  As $\tau$ becomes larger, this assumption begins to break down as the wells become ``shallow'' \cite{hanggi1990reaction}.  We expand $W(\theta)$ around the minimum $\theta = \pi/2$, defining $\delta = \theta-\pi/2$, $W(\delta) \approx -\frac{\pi^2}{8} + \frac{1}{2} \delta^2$.  This shows that (neglecting transitions between the two states) the distribution of $\delta$, $P(\delta) \sim e^{-W(\delta)/\tau}$, is a Gaussian, centered around zero, with a standard deviation of $\tau$.  When $\tau \sim \pi/2$, our idealized picture of hopping between wells breaks down.  

In the strong-confinement limit $\kappa \gg 1$, the only controlling parameter for our model is $\tau$.  Can we estimate $\tau$ from experimental data?  To determine $\tau$ from \eq \ref{eq:mfpt_exact}, we need to know 1) the rate of spontaneous reversals, and 2) the persistence time $P$ that sets our unit of time.  Desai et al. \cite{desai2013contact} have recently measured the rate of spontaneous reversal of NRK-52E rat epithelial kidney cells on micropatterned adhesive substrates, finding a rate of $0.08 \pm 0.05$ reversals per hour.  The characteristic persistence time $P$ for this system has not been measured, but we can estimate it.  For a single cell without confinement ($\kappa = 0$), the alignment term of our model vanishes, as $\hat{\vb{v}} = \hat{\vb{p}}$, and the Szabo model reduces to a generic self-propelled particle model with velocity-velocity correlation function (in dimensional units), $\langle \vb{\tilde{v}}(\tilde{t})\cdot\vb{\tilde{v}}(0) \rangle = v_0^2 e^{-\tilde{t}/P}$ \cite{peruani2007self,campos2010persistent}.  This form and the corresponding result for the mean-squared displacement have historically been used to analyze cell motion \cite{gail1970locomotion,selmeczi2005cell,stokes1991migration,hartman1994fundamental}.  Persistence times range from tens of seconds (for neutrophils \cite{hartman1994fundamental}) to the order of hours (fibroblasts \cite{gail1970locomotion} and endothelial cells \cite{stokes1991migration}).  Based on this information, and the trajectories of unconfined NRK-52E cells shown in \cite{desai2013contact}, we estimate $P\approx 1$ hour.  Using this estimate, we fit to $k_{\textrm{reversal}} = (2 \tm P)^{-1}$, where $\tm$ is given by \eq \ref{eq:mfpt_exact} \cite{footnote_factor_of_two}.  We find $\tau \approx 0.3$ (i.e. $T \approx 0.3$ hr).  This is a measurement of the cell's internal memory as well as the strength of cell-cell alignment in collective motility.  

\begin{figure}[ht!]
\includegraphics[width=90mm]{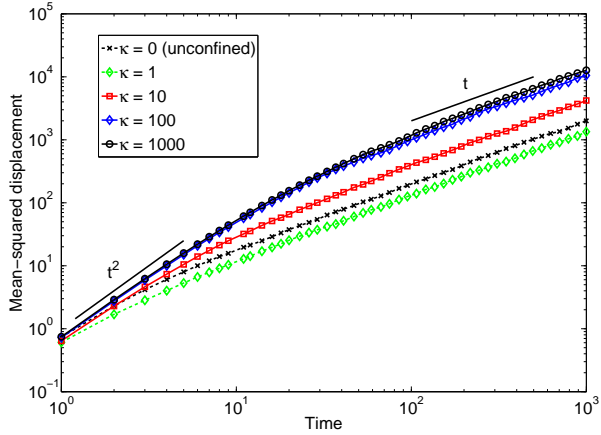}
\caption{Unitless mean-squared displacements $\langle |\Delta \vb{r}|^2 \rangle$ of a cell from simulation of \eqs \ref{eq:r}-\ref{eq:theta} with $\tau = 0.3$ and varying values of $\kappa$.  The initial increase of $\kappa$ from zero to one decreases the dispersal of the cell, due to its confinement.  However, for larger values of $\kappa$, the confinement-induced persistence ensures that the cell persists in ballistic motion for a longer time before reorienting, increasing the cell dispersal.}
\label{fig:msd}
\end{figure}

How does the increased persistence time alter the cell's dispersal?  For an unconfined cell, the mean-squared displacement increases ballistically as $\langle |\Delta \vb{r}|^2 \rangle \sim t^2$ at short times.  At longer times, reorientation leads to a diffusive motion, $\langle |\Delta \vb{r}|^2 \rangle \sim t$ \cite{peruani2007self}.  We show in \fig \ref{fig:msd} the mean-squared displacement of a cell under increasing confinement $\kappa$.  Interestingly, increasing $\kappa$ changes cell dispersal non-monotonically.  The presence of a nonzero $\kappa$ significantly decreases cell displacement ($\kappa = 1$ in \fig \ref{fig:msd}): confinement prevents the cell from traveling beyond $x = 1/\kappa$.  As $\kappa$ is increased, the cell's persistence increases markedly, as studied above.  The cell then maintains a steady crawling motion for much longer without reorienting, leading to an extended period of ballistic motion and a larger dispersal (\fig \ref{fig:msd}).  This mean-squared displacement begins to saturate as we reach the strong confinement limit for $\kappa > 100$.

\section{Collective motility under confinement}

\begin{figure}[ht!]
\includegraphics[width=90mm]{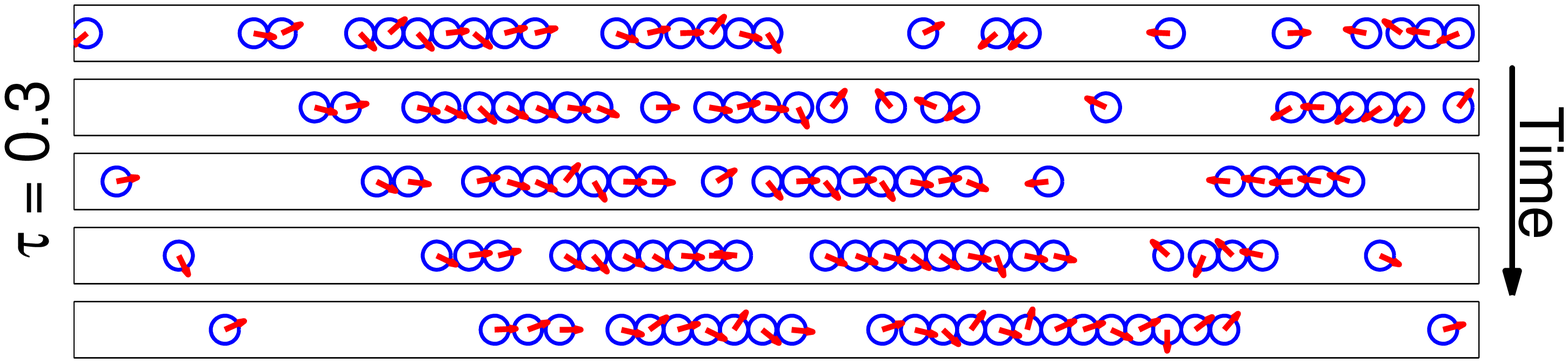}
\includegraphics[width=90mm]{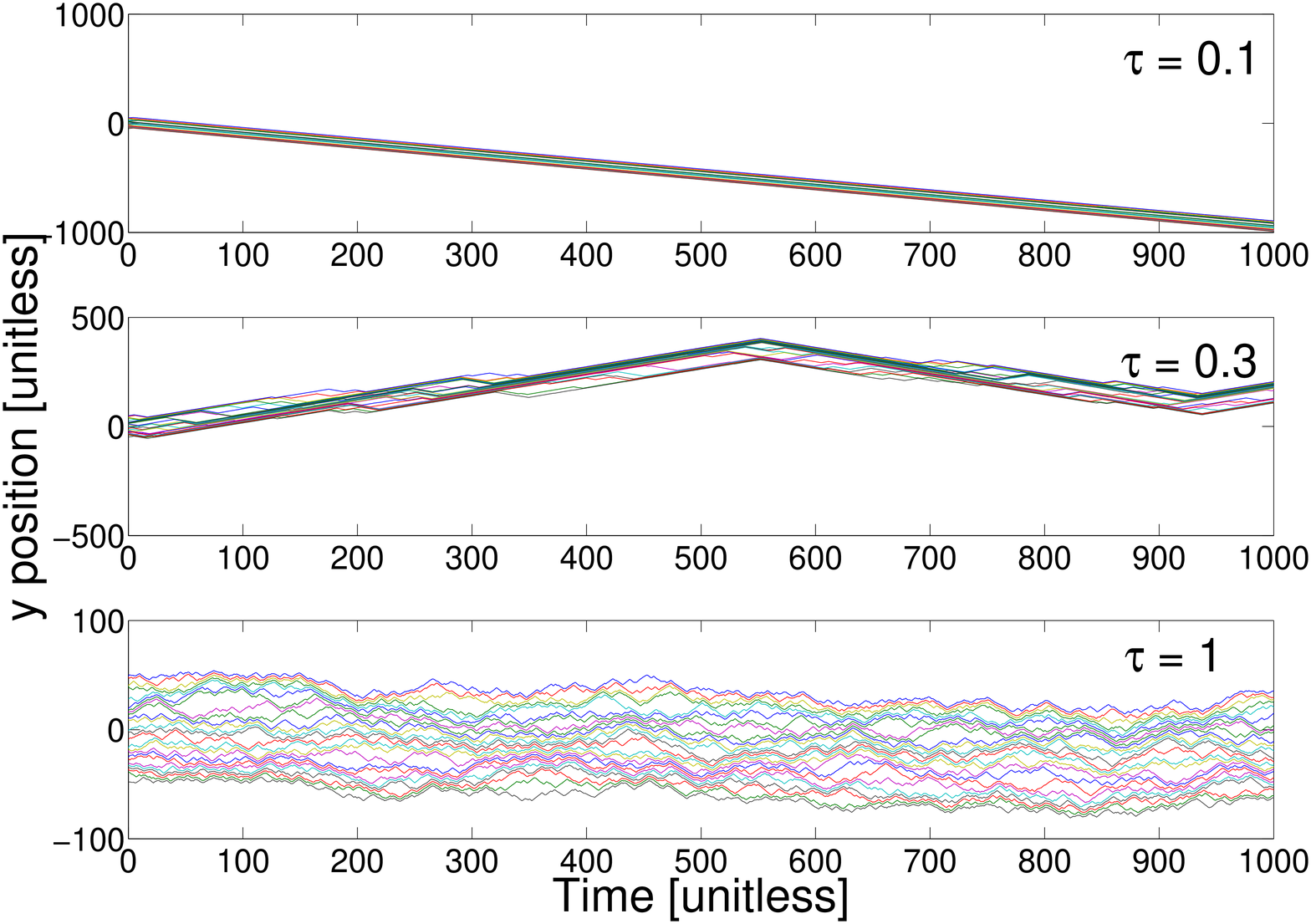}
\caption{Velocity-aligning cells with cell-cell repulsion develop ``trains'' consistent with \cite{desai2013contact}.  TOP: representative snapshots of trains of cells when $\tau = 0.3$; circles represent cell size while arrows represent cell polarity $\hat{\vb{p}}$.  Each snapshot is separated by a unitless time of 5 (e.g. 5 hours if $P = 1$ hour).  BOTTOM: Positions of cells are plotted over time for $\tau = 0.1, 0.3, 1$.  Trains are highly persistent for $\tau = 0.1$, occasionally reverse for $\tau = 0.3$, and are only transient if $\tau = 1$.  In this simulation, 25 cells of radius $R = 1$ are confined in a periodic system of length $L = 100$.  The confinement strength $\kappa = 1000$, and the cell-cell repulsion is $\kappa_{\textrm{cell}} = 100$.}
\label{fig:flock}
\end{figure}

Collective motion emerges in a simple extension of our model to multiple cells in confinement.  We simulate multiple velocity-aligning cells, interacting only by a short-range repulsion force. In unitless variables, the force between cells $i$ and $j$ is $\vb{F}_{ij} = -\kappa_{\textrm{cell}} (2R-|\vb{r}_{ij}|)\Theta(2R-|\vb{r}_{ij}|) \hat{\vb{r}}_{ij}$, where $\vb{r}_{ij} = \vb{r}_j - \vb{r}_i$, $R$ the cell radius, and $\Theta$ is the Heaviside step function.  As noted by Henkes et al., \cite{henkes2011active}, simple repulsive interactions are sufficient to generate collective motion.  We find trains of cells as observed in Desai et al. \cite{desai2013contact} (\fig \ref{fig:flock}).  For $\tau = 0.1$, all cells quickly align into a single direction, and very rarely collectively reverse (observed once in a simulation of unitless time $10^4$).  For the experimentally estimated value of $\tau = 0.3$, we see occasional reversals of trains, generally consistent with the results of \cite{desai2013contact}.  However, for $\tau = 1$, no persistent collective motions occur and trains are transient.  These results are similar to the observations of \cite{desai2013contact}, but have an important caveat.  Our simulations predict that cells reverse at a physical barrier; those studied in \cite{desai2013contact} do not.  Our model may therefore be more appropriate for fibroblasts as studied in \cite{doyleyamada}, which are observed to reverse at micropattern ends; we primarily study the results of Desai et al. because they have quantified the spontaneous repolarization rate.  

Our results show that in a velocity-aligning model, the increase in persistence time of a strongly confined cell and the interactions driving collective cell motility are intimately related.  Balzer et al. have recently shown that human breast carcinoma (MDA-MB-231) is highly persistent in a confined channel, but that this persistence can be disrupted by interfering with microtubule polymerization or depolymerization by application of colchicine or paclitaxel (Taxol) \cite{balzer2012physical}.  This raises an interesting question: do Taxol and colchicine disrupt collective cell migration?  Our model combined with the data of Balzer et al. \cite{balzer2012physical} suggests that Taxol and colchicine would have similar effects on collective migration - even though they have opposing effects on the stability of microtubules.  The idea that velocity alignment may be linked with microtubule dynamics is perhaps not surprising, given the known roles of microtubules in cell polarity \cite{sugioka2012formation}.  

While we have worked with the simplest possible model, our results may be extended to more complex cellular dynamics and potentially used to relate single-cell behavior under strong confinement with cell-cell interactions.  To do this, we may have to extend this model.  Experimental cell tracks show velocity-velocity correlations with two distinct time scales, rather than single-exponential as assumed here \cite{selmeczi2005cell}; this feature may be added by adding a stochastic process controlling the cell speed, i.e. $v_0\to v(t)$ \cite{peruani2007self}.  In our model, $v_0$ only rescales the lengths involved; if $v(t)$ does not frequently drop to zero, we expect the varying velocity to only affect our results minimally.  We argue that the confining potential's details are relatively unimportant; simulations with strong hard-wall confinement are consistent with \eq \ref{eq:mfpt} (Appendix \ref{app:hardwall}).  Detailed cellular simulations show adhesion to extracellular matrix also increases persistence \cite{szabo2012invasion}; our results may help explain this.  

In this paper, we have shown that even in a very simple model, cell motility in confinement can take on a profoundly different character than on a two-dimensional substrate, without invoking different mechanisms for free- and confined- cell motility.  There may, of course, be other reasons to argue for biophysical differences between free and confined motility \cite{hawkins2009pushing,friedl2003tumour}.  Our results may explain the origin of large persistence times experimentally observed for confined cells.  We believe that these results are useful as a baseline model for the analysis of cell crawling in confinement, as well as for making connections between single- and collective- cell motility.  In particular, our technique provides an in principle straightforward way to determine the velocity-alignment timescale that is important for collective cell motion \cite{henkes2011active,basan2013alignment,szabo2006phase,kabla2012collective,szabo2010collective} by the analysis of single-cell trajectories under strong confinement.  This allows an interesting test of these minimal models of collective cell motions.  

\section*{Acknowledgments}

This work was supported by NIH Grant No. P01 GM078586, NSF Grant No. DMS 1309542, and by the Center for Theoretical Biological Physics.  BAC acknowledges helpful discussions with Yaouen Fily, Christopher Pierse, and Yaojun Zhang.  

\appendix

\section{Determining turnaround times}
\label{app:turnaround}

We are interested in characterizing the typical time required for the cell to orient itself from the $\hat{\vb{y}}$ direction to the $-\hat{\vb{y}}$ direction.  This value should be directly comparable to the mean-first-passage-time (MFPT) result derived in the main paper.  In the MFPT result, we calculate the rate for the cell's orientation to move from $\theta = \pi/2$ to $\theta = 0$ or $\theta = \pi$, and assume that $\theta = \pi/2$ and $\theta = \pi$ are absorbing boundaries.  By symmetry, this will be precisely {\it twice} the actual rate, because a cell with orientation $\theta = 0$ is equally likely to transition to the ``potential wells'' at $\theta = \pi/2$ and $\theta = -\pi/2$.  

\begin{figure}[ht!]
\includegraphics[width=95mm]{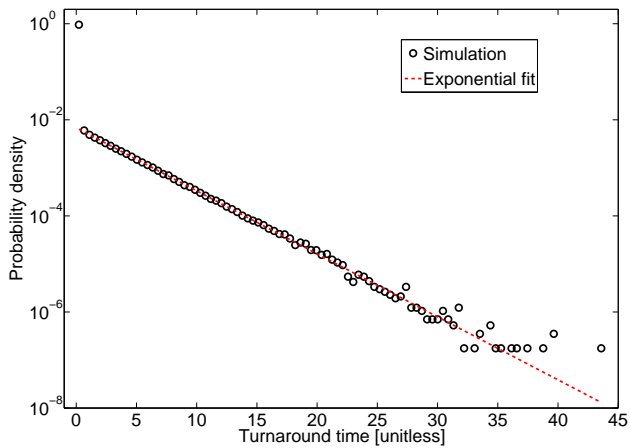}
\caption{Distribution of turnaround times computed for simulation with $\kappa = 1000$ and $\tau = 0.5$.  $\Delta t = 10^{-4}$, total simulation time $10^6$ (unitless).  }
\label{fig:histogram}
\end{figure}

In our simulations, it is convenient to track these transitions with a similar absorbing boundary condition assumption.  In order to do this, we track the times between turnaround events, i.e. events where $\sin \theta$ changes sign.  A histogram of these events is shown in \fig \ref{fig:histogram}.  If the transition between directions is characterized by a simple rate, we would expect an exponential distribution of transition times.  We see that the tail of this distribution is very well fit by an exponential; however, there is a peak at small times.  This peak arises from events where the cell's orientation remains close to $\sin \theta = 0$, and mostly depends on the time step we use.  We therefore fit to the exponential tail to find the turnaround time, as shown in \fig \ref{fig:histogram}.

\section{Turnaround times under hard wall confinement}
\label{app:hardwall}

Our results for the turnaround time in the limit $\kappa \gg 1$ do not depend on the strength of confinement $\kappa$; we would therefore expect them to apply for strong confinement in other potentials.  Here, we show that this is true for a velocity-aligning particle confined by hard walls at $x = \pm w/2$.  In this case, the strong confinement limit occurs when $w \ll 1$, i.e. the cell does not reorient quickly before it impacts a wall; this strong confinement limit has been recently explored by Fily et al. \cite{fily2014dynamics} for many confining geometries, though in the absence of velocity alignment.  

Our model for hard wall confinement is 
\begin{align}
\label{eq:rhw}
\partial_t \vb{r} &= \vb{\hat{p}} + \vb{F}_{\textrm{wall}} \\
\partial_t \theta &= \frac{1}{\tau} \textrm{arcsin}\left[ \left(\vb{\hat{p}}\times\vb{\hat{v}}\right)_z \right] + \xi(t) \label{eq:thetahw}.
\end{align}
where, as in \cite{fily2014dynamics}, $\vb{F}_{\textrm{wall}} = -\hat{p}_x \hat{\vb{x}} = -\cos\theta \hat{\vb{x}}$ when the particle is on the wall ($x = \pm w/2$) and the polarity is pointing toward the wall, and zero otherwise, i.e. the wall exerts a force sufficient to keep the particle from penetrating it.  In practice, we evolve this with an adaptive step algorithm.  This takes the form:

\begin{enumerate}
\item Attempt to evolve Eqs. \ref{eq:rhw}-\ref{eq:thetahw} forward by $\Delta t$.  Here, and throughout the paper, we use the simplest possible Euler-Maruyama method to integrate our equations of motion \cite{kloeden1992numerical}.
\item If the new position crosses $x = \pm w/2$, solve for the time $\alpha$ at which this occurs, $\alpha = \left[\pm w/2 - x(t)\right]/v_x$.
\begin{enumerate}
\item Evolve Eqs. \ref{eq:rhw}-\ref{eq:thetahw} forward by $\alpha$
\item Set $\vb{F}_{\textrm{wall}} = -\hat{p}_x \hat{\vb{x}}$ and therefore $v_x = 0$
\item Evolve Eqs. \ref{eq:rhw}-\ref{eq:thetahw} forward by $\Delta t - \alpha$
\end{enumerate}
\end{enumerate}

\begin{figure}[t!]
\includegraphics[width=100mm]{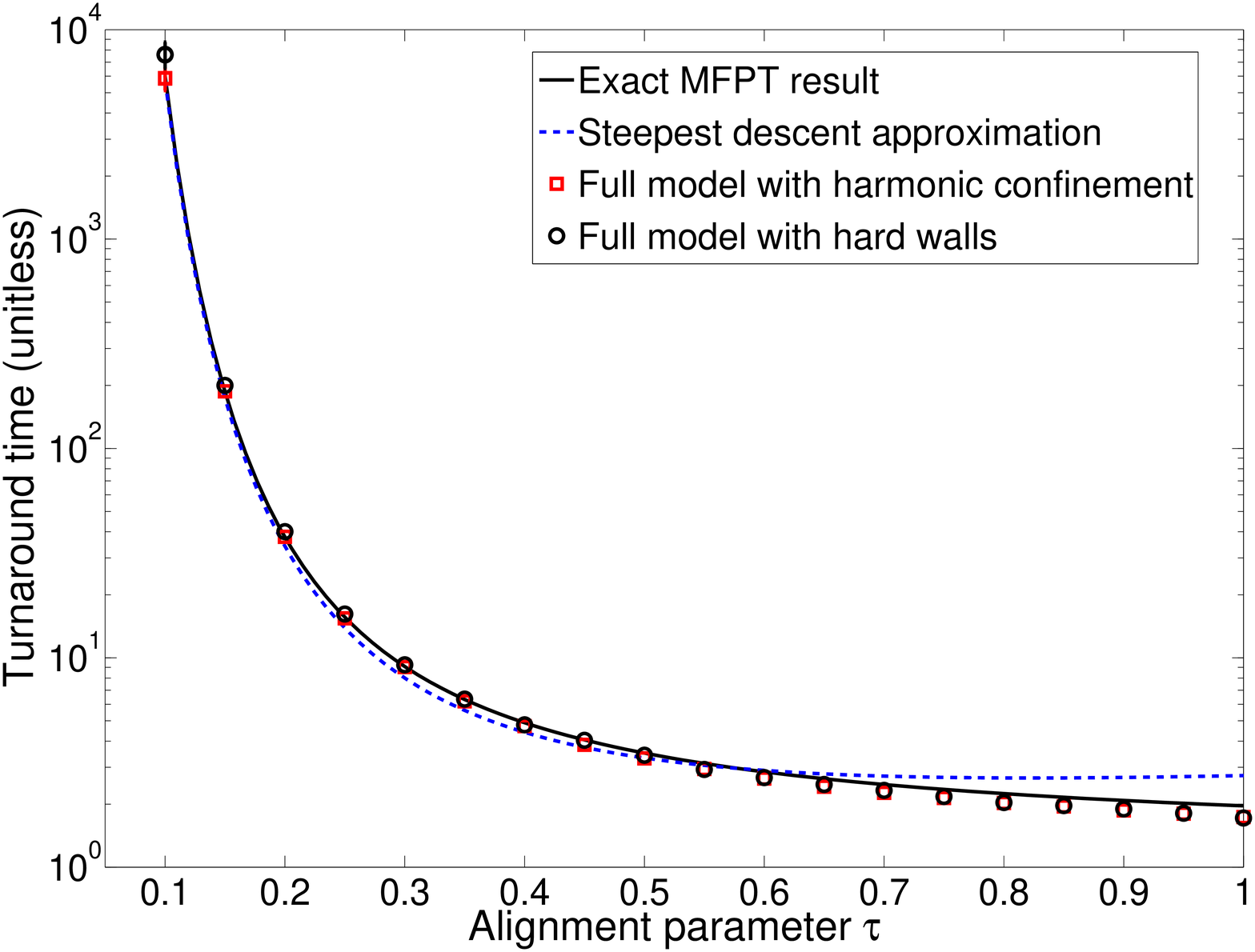}
\caption{Turnaround times for strongly-confined velocity aligning particles under harmonic and hard wall confinement are the same.  Harmonic confinement simulations performed with $\kappa = 1000$, $\Delta t = 10^{-4}$, and a simulation time of $10^6$; hard wall simulations performed with $w = 4 \times 10^{-4}$, $\Delta t = 10^{-4}$ and simulation times of $10^6$ (for $\tau = 0.1-0.25$) or $10^5$ (for $\tau = 0.3$ and larger).  Error bars are computed by the bootstrap method; see text.}  
\label{fig:rateshw}
\end{figure}

We show in \fig \ref{fig:rateshw} that the turnaround times are consistent with those from our harmonic confinement simulations presented in the main paper.  Error bars in \fig \ref{fig:rateshw} are computed by applying the bootstrap method \cite{efron1982jackknife} to the fitting approach shown above; error bars for $\tau > 0.1$ are on the order of symbol size or smaller, and are not shown.  

\begin{thebibliography}{46}
\bibitem [{foo()}]{footnote_factor_of_two}%
  \BibitemOpen
  \href@noop {} {}\bibinfo {note} {In calculating the mean first passage time
  in both theory and our simulations, we have assumed that the peaks of
  $W(\theta)$ are absorbing barriers. In fact, by symmetry only half of
  trajectories that reach a peak will transition into the adjacent well. To
  calculate the true rate, we must divide the mean-first-passage rate by two,
  e.g. $k_{\textrm{reversal}} = (2 \tm P)^{-1}$ (in real units).}\BibitemShut
  {Stop}%
\end{thebibliography}

%

\end{document}